\documentclass[conference]{IEEEtran}
\IEEEoverridecommandlockouts
\usepackage{cite}
\usepackage{amsmath,amssymb,amsfonts}
\usepackage{algorithmic}
\usepackage{graphicx}
\usepackage{textcomp}
\usepackage{xcolor}
\usepackage{subcaption}
\usepackage{float}
\newcommand{\commentout}[1]{}
\def\BibTeX{{\rm B\kern-.05em{\sc i\kern-.025em b}\kern-.08em
    T\kern-.1667em\lower.7ex\hbox{E}\kern-.125emX}}
\begin{document}

\title{Tiled Beamspace MVDR for 1024-element Wideband Radar \\
}

\author{\IEEEauthorblockN{Oveys Delafrooz Noroozi}
\IEEEauthorblockA{
\textit{University of California, Santa Barbara}\\
Santa Barbara, CA, U.S.A. \\
oveys@ucsb.edu}
\and
\IEEEauthorblockN{Jiyoon Han}
\IEEEauthorblockA{
\textit{University of Michigan}\\
Ann Arbor, MI, U.S.A. \\
hanjyoon@umich.edu}
\and
\IEEEauthorblockN{Wei Tang}
\IEEEauthorblockA{
\textit{University of Michigan}\\
Ann Arbor, MI, U.S.A. \\
weitang@umich.edu}
\and \hspace{3cm}
\IEEEauthorblockN{Zhengya Zhang}
\IEEEauthorblockA{\hspace{3cm}
\textit{University of Michigan}\\ \hspace{3cm}
Ann Arbor, MI, U.S.A. \\ \hspace{3cm}
zhengya@umich.edu}
\and
\IEEEauthorblockN{Upamanyu Madhow}
\IEEEauthorblockA{
\textit{University of California, Santa Barbara}\\
Santa Barbara, CA, U.S.A. \\
madhow@ucsb.edu}
}

\maketitle

\begin{abstract}
We present a tiled architecture for computationally efficient digital beamforming for wideband massive MIMO radar, using beamspace dimension reduction for each tile, and coordinated training of reduced-dimension MVDR beamformers across tiles.  We illustrate the efficacy of our approach for a setting in which a 1024-element airborne radar platform beamforms towards airborne targets while suppressing strong interference from ground transmitters.   The array is organized into eight 128-element tiles, each a 2D array with $4 {\rm ~(vertical)} \times 32 {\rm ~(horizontal)} $ elements.  Each tile applies a 2D spatial DFT to achieve energy concentration in beamspace, and a 1D temporal FFT to channelize the wideband signal into subbands for which narrowband array models apply.  A small tile-level beamspace window is selected for each target (depending on its angle of arrival) in each subband, and coordinated training across tiles is used to compute reduced-dimension MVDR beamformers per-target, per-subband.  While full-dimensional MVDR processing is infeasible for the system under consideration, we show that our proposed approach significantly outperforms beamspace MVDR beamforming for a single 128-element tile, where we set the dimensions of the spatial filter (and hence the complexity of MVDR training) to be equal in both systems.
\end{abstract}

\section{Introduction}
\label{Sec:Introduction}
Fully digital wideband MIMO radar arrays offer the ability to form multiple beams simultaneously for target detection, tracking, and interference suppression, in contrast to the sequential scanning beams of traditional phased arrays. Continued advances in RF integration and digital processing are rapidly making massive digital arrays—featuring hundreds or thousands of elements—feasible for next-generation radar systems. However, the computational burden associated with conventional spatial processing presents a fundamental obstacle to such scaling. Classical minimum variance distortionless response (MVDR) beamforming requires the estimation and inversion of an $N \times N$ sample covariance matrix, leading to $O(N^3)$ complexity, which becomes prohibitive for large arrays.

Beamspace dimension reduction is a promising pathway to overcoming this limitation. For regularly spaced arrays, the spatial signature of a plane wave is a complex exponential, enabling its energy to be concentrated by a spatial FFT. Beamspace concepts have long been explored to improve robustness and reduce computational load across a broad range of radar and array-processing applications, including through-the-wall radar imaging \cite{yoon2008high, yoon2011mvdr}, weather radar \cite{nai2016adaptive},
moving-target localization \cite{jin2002beamspace, wang2008mvdr}, and subarray-based interference mitigation \cite{doisy2010interference}. Early work on partial adaptation and beamspace MUSIC also demonstrated the benefits of dimension reduction for high-resolution
applications \cite{zoltowski1990simultaneous}. More recently, beamspace dimension reduction has been successfully employed for reduced-complexity multiuser MIMO reception \cite{abdelghany2019beamspace, abdelghany2020scalable}, with validation via channel models derived from measured data \cite{cebeci2024scaling}. These developments motivate the use of beamspace techniques as an enabling mechanism for scalable adaptive all-digital processing in massive arrays for radar.

Our earlier work \cite{noroozi2025scaling} demonstrated that energy concentration in beamspace enables wideband
windowed-beamspace MVDR processing with complexity scaling dominated by the $O(N \log N)$ cost of spatial FFTs. However,
further architectural innovations are required to support \emph{spatially massive} arrays whose total elements exceed what can be processed as a single monolithic block. Recent work in multiuser MIMO reception shows the promise of tiled beamspace dimension reduction \cite{han2024tiled}, and motivates the approach presented in this paper.

In this paper, we extend reduced-dimension beamspace MVDR beamforming for wideband radar to a tiled architecture that enables scaling to massive antenna arrays by parallelizing computations across tiles, across subbands, and across targets. Each tile performs a local 2D spatial FFT followed by an angle-of-arrival (AoA) dependent beamspace window that extracts, for each subband, the dominant DFT bins for a desired beamforming direction corresponding to each target being tracked. The windowed outputs from all tiles are concatenated for covariance estimation and adaptation of a reduced-dimensional MVDR beamformer.  Application of the resulting MVDR beamformers is parallelized across tiles with minimal requirements for inter-tile communication.  Our performance evaluation considers a 1024-element array, organized in eight 128-element tiles, which is representative of the scale envisioned for emerging massive MIMO digital radar.

\section{System Model}
\label{Sec:System Model}
We consider a wideband radar system employing a uniform planar array (UPA) partitioned into identical tiles to enable scalable processing. The array is divided into $T_z \times T_x$ tiles, resulting in a total of $T = T_z T_x$ tiles. Each tile contains $N_z$ vertical and $N_x$ horizontal antenna elements, for a total of $N = N_z N_x$ antennas per tile. All tiles share a common transmit waveform and are synchronized in both time and frequency. The array elements are uniformly spaced with inter-element spacing $d = \lambda/2$, where $\lambda$ is the wavelength corresponding to the design frequency $f_d$.

The azimuth and elevation angles of a $k$-th received signal are denoted by $(\varphi_k, \theta_k)$. The corresponding reference spatial frequencies at the design frequency $f_d$ are
\begin{equation}
    \boldsymbol{\Omega}_k^{\mathrm{ref}}
    = \pi
    \begin{bmatrix}
        \cos\theta_k \sin\varphi_k \\[1pt]
        \sin\theta_k
    \end{bmatrix}.
    \label{eq:omega_ref}
\end{equation}
At any processing frequency $f$, the spatial frequencies scale linearly as
\begin{equation}
    \boldsymbol{\Omega}_k(f) =
        \begin{bmatrix}
        \Omega_{k,x}(f) \\[1pt]
        \Omega_{k,z}(f)
    \end{bmatrix} = \frac{f}{f_d} \, \boldsymbol{\Omega}_k^{\mathrm{ref}}.
    \label{eq:omega_scale}
\end{equation}

For a uniform linear array of $N$ elements and spatial frequency $\Omega$, the steering vector is defined as
\begin{equation}
    \mathbf{u}_N(\Omega)
    =
    \begin{bmatrix}
        1 & e^{j\Omega} & \cdots & e^{j(N-1)\Omega}
    \end{bmatrix}^{\!\top}.
    \label{eq:aN}
\end{equation}
Accordingly, $\boldsymbol{\psi}_k(f) \in \mathbb{C}^N$ is the in-tile (element-level) array response for the $k$-th source
\begin{equation}
    \boldsymbol{\psi}_k(f)
    =
    \mathbf{u}_{N_x}\!\big(\Omega_{k,x}(f)\big)
    \otimes
    \mathbf{u}_{N_z}\!\big(\Omega_{k,z}(f)\big),
    \label{eq:psi_elem}
\end{equation}

The tile-level array response is defined using the same construction applied to the tile grid:
\begin{equation}
    \boldsymbol{\Psi}_k(f)
    =
    \mathbf{u}_{T_x}\!\big(N_x\Omega_{k,x}(f)\big)
    \otimes
    \mathbf{u}_{T_z}\!\big(N_z\Omega_{k,z}(f)\big),
    \label{eq:phi_tile}
\end{equation}
where $\boldsymbol{\Psi}_k(f) \in \mathbb{C}^T$ describes the spatial phase progression across tiles.

The per-tile steering vector observed by tile $t$ is therefore expressed as
\begin{equation}
    \mathbf{a}_k^{(t)}(f) =
    [\boldsymbol{\Psi}_k(f)]_t \boldsymbol{\psi}_k(f),
    \quad t = 1, \dots, T.
    \label{eq:per_tile_channel}
\end{equation}

The received snapshot at a given subband and tile $t$ is modeled as
\begin{equation}
    \mathbf{y}^{(t)}[n]
    =
    \sum_{k=1}^{K}
    \alpha^{(t)}_k\mathbf{a}_k^{(t)} \, p_k[n - \tau_k]
    + \mathbf{I}^{(t)}[n]
    + \mathbf{n}^{(t)}[n],
    \label{eq:per_tile_y}
\end{equation}
where
\begin{itemize}
    \item $\alpha^{(t)}_k \in \mathbb{C}$ is the complex gain of the $k$-th source,
    \item $p_k[\cdot]$ is the transmit pulse delayed by $\tau_k$,
    \item $\mathbf{I}^{(t)}[n]$ models interference or clutter at tile $t$,
    \item $\mathbf{n}^{(t)}[n] \sim \mathcal{CN}(\mathbf{0}, \sigma_t^2 \mathbf{I})$ denotes spatially white noise.
\end{itemize}
The frequency (subband) index is omitted for notational simplicity, as the signal model is identical across all subbands.

To simplify notation when combining tile-level quantities, we introduce the concatenation operator
\begin{equation}
    \left\{\mathbf{x}^{(t)}\right\}_{t=1}^{T}
    \triangleq
    \begin{bmatrix}
        \mathbf{x}^{(1)\top} \;
        \mathbf{x}^{(2)\top} \;
        \cdots \;
        \mathbf{x}^{(T)\top}
    \end{bmatrix}^{\!\top},
    \label{eq:cat_def}
\end{equation}
which stacks a collection of tile-specific vectors into a single global vector.

Using this operator, the received snapshot at time $n$ across all tiles can be compactly written as
\begin{equation}
    \mathbf{y}[n]
    =
    \left\{\,\mathbf{y}^{(t)}[n]\,\right\}_{t=1}^{T}
    \in \mathbb{C}^{TN}.
    \label{eq:stacked_y}
\end{equation}

\begin{figure}[t]
    \centering
    \includegraphics[width=0.8\linewidth, height=0.5\linewidth]{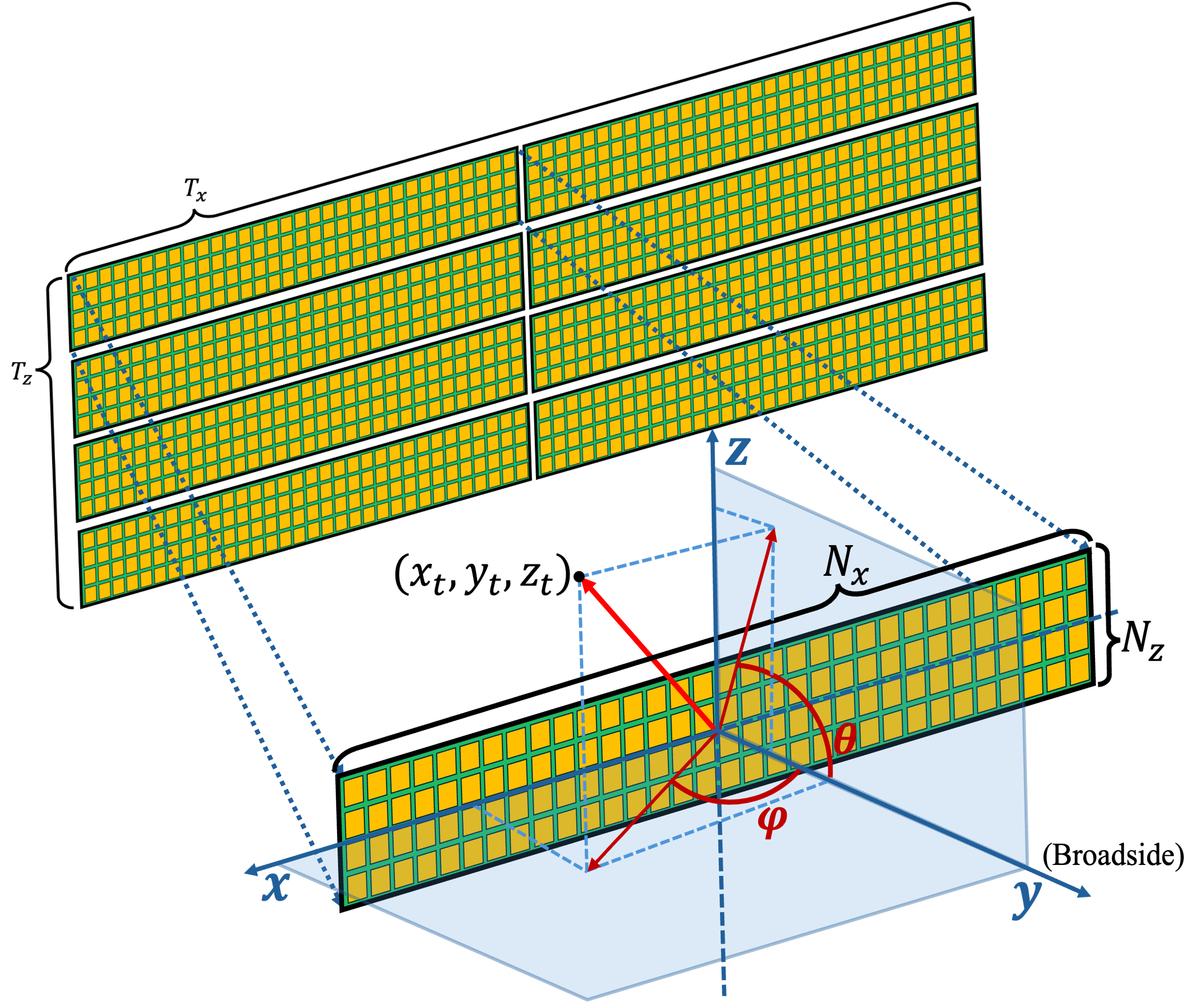}
        \caption{\small
        Tiled UPA geometry illustrating the $T_z \times T_x$ tile partitioning, 
        the per-tile subarray structure with $N_z \times N_x$ elements, and the definitions of 
        elevation $(\theta)$ and azimuth $(\varphi)$ angles corresponding to the target location \((x_t, y_t, z_t)\).
        }
    \label{Fig:Geometry}
\end{figure}

We briefly recall the well-known MVDR beamformer. For each subband and target index $k$, the corresponding correlator vector $\mathbf{c}_k$ is given by
\begin{equation}
    \mathbf{c}_k =
    \frac{\widehat{\mathbf{R}}^{-1}\mathbf{a}_k}
    {\mathbf{a}_k^H \widehat{\mathbf{R}}^{-1}\mathbf{a}_k}.
    \label{Eq:CorrMVDR}
\end{equation}
where $\widehat{\mathbf{R}}$ denotes the sample covariance matrix of the received snapshots $\{ \mathbf{y}[n] \}$,
\begin{equation}
    \widehat{\mathbf{R}} = \frac{1}{n_t}\sum_{n=1}^{n_t}\mathbf{y}[n]\mathbf{y}^H[n],
    \label{Eq:EmpiricalCov}
\end{equation}
and $\mathbf{a}_k$ is the global steering vector of the $k$-th source obtained by concatenating the per-tile steering vectors,
\begin{equation}
    \mathbf{a}_k =
    \left\{\mathbf{a}_k^{(t)}\right\}_{t=1}^{T}
    \in \mathbb{C}^{TN}.
    \label{eq:stacked_y}
\end{equation}
Applying the correlator $\mathbf{c}_k$ to the received vector $\mathbf{y}[n]$ (i.e., $\mathbf{c}_k^H\mathbf{y}[n]$) produces the beamformer output for the $k$-th source direction.

The preceding operations become computationally infeasible for large $N$ (formation of $\widehat{\mathbf{R}}$ is $O(N^3)$ while inversion is around $O(N^{2.5})$), motivating the dimension-reduction strategies considered here.  

\section{Tiled Beamspace Architecture}
\label{Sec:Proposed System}
\begin{figure*}[t]
    \centering
    \includegraphics[width=0.75\textwidth, height=0.35\textwidth]{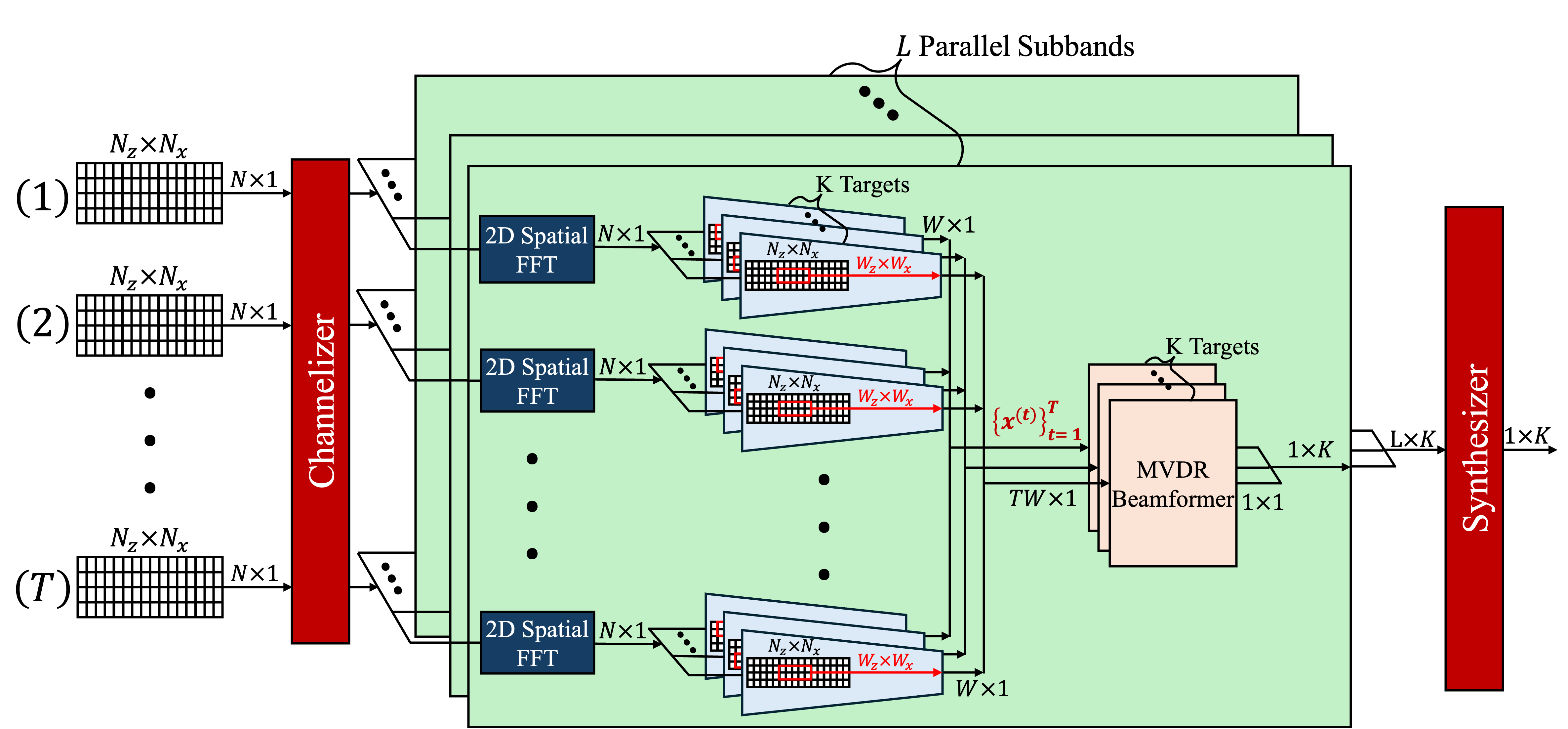}
    \caption{\small
    Coordinated tiled beamspace MVDR processing for wideband radar signals. After channelization, each tile performs 2D spatial FFT projection and AoA-dependent beamspace windowing to produce $W \times 1$ reduced-dimension vectors per target. The $T$ tile outputs are concatenated into a global $TW \times 1$ vector that feeds a centralized MVDR beamformer, followed by synthesis of a wideband signal for standard range-Doppler processing.
    }
    \label{Fig:SystemModel}
\end{figure*}

This section describes the proposed tiled beamspace architecture for scalable wideband radar arrays. The array is partitioned into $T$ identical tiles, each containing an $N_z \times N_x$ subarray of antenna elements. Figure~\ref{Fig:SystemModel} illustrates the processing flow of the coordinated tiled windowed–beamspace MVDR architecture.

After channelization, the received signal is decomposed into $L$ parallel subbands. In each subband, every tile independently applies a 2D spatial DFT to obtain a beamspace representation of the received snapshot. An AoA-dependent window selects a fixed $W_z \times W_x$ region of the beamspace output, yielding a reduced-dimension $W \times 1$ vector per tile.

The windowed outputs from all $T$ tiles are concatenated into a global $TW \times 1$ beamspace snapshot. During training, these reduced-dimension observations are used to estimate the covariance matrix and compute MVDR beamforming weights for each target and subband. Although Fig.~\ref{Fig:SystemModel} illustrates this computation in a central processing unit, the resulting weights are distributed back to the tiles and applied locally during inference, enabling parallel beamforming across the array. The MVDR outputs for the $K$ targets are then passed to the wideband synthesizer, which recombines the $L$ subbands to produce the appropriate wideband signal for final range-Doppler processing.

\subsection{Beamspace Transformation and Windowing}\label{Sec:BeamWindow}

To enable scalable spatial processing in the tiled architecture, each tile projects its received signal into beamspace using a two-dimensional discrete Fourier transform. Let $\mathbf{D}_N \in \mathbb{C}^{N \times N}$ denote the unitary $N$-point DFT matrix with entries
\begin{equation}\nonumber
    [\mathbf{D}_N]_{m,n}
    =
    \frac{1}{\sqrt{N}} e^{-j \frac{2\pi}{N} (m-1)(n-1)},
    \qquad m,n = 1,\dots,N.
    \label{eq:DN_def}
\end{equation}
The beamspace transform for a tile with $N_z$ vertical and $N_x$ horizontal antenna elements is given by the Kronecker product
\begin{equation}\nonumber
    \mathbf{D}
    =
    \mathbf{D}_{N_x}^{\top} \otimes \mathbf{D}_{N_z}
    \in \mathbb{C}^{N \times N},
    \qquad N = N_z N_x.
    \label{eq:D_kron}
\end{equation}

To reduce dimensionality, we apply an AoA-dependent beamspace window that extracts the spatial DFT bins associated with the $k$-th target.  
Let $\mathbf{S}^{(k)}_{W_z} \in \mathbb{R}^{W_z \times N_z}$ and 
$\mathbf{S}^{(k)}_{W_x} \in \mathbb{R}^{W_x \times N_x}$ denote the vertical and horizontal
binary selector matrices that extract $W_z$ and $W_x$ beamspace components, respectively, centered around the DFT location corresponding to the $k$-th target’s AoA.  
This selects a total of $W = W_z W_x$ beamspace coefficients from the original $N = N_z N_x$ samples.

The full 2D windowing operator for the $k$-th target is constructed as
\begin{equation} \nonumber
    \mathbf{S}_k
    =
    \mathbf{S}^{(k)\top}_{W_x}
    \otimes
    \mathbf{S}^{(k)}_{W_z}
    \in \mathbb{R}^{W \times N}.
    \label{eq:window_operator_target}
\end{equation}

For tile $t$, the corresponding beamspace dimension-reduced snapshot of the $k$th target is

\begin{equation}
    \widetilde{\mathbf{y}}_k^{(t)}[n]
    =
    \mathbf{S}_k\mathbf{D} \, \mathbf{y}^{(t)}[n]
    \in \mathbb{C}^{W}.
    \label{eq:tile_beamspace}
\end{equation}

Finally, the global windowed beamspace snapshot across all tiles is obtained using the concatenation operator defined in \eqref{eq:cat_def}:
\begin{equation}
    \widetilde{\mathbf{y}}_k[n]
    =
    \left\{\widetilde{\mathbf{y}}^{(t)}_k[n]\right\}_{t=1}^{T}
    \in \mathbb{C}^{TW}.
    \label{eq:global_tile_windowed}
\end{equation}

\subsection{Reduced-Dimension MVDR}\label{Sec:BeamMVDR}

Given the dimension-reduced signal $ \widetilde{\mathbf{y}}_k[n] \in \mathbb{C}^{TW}$, we compute the beamspace MVDR weights using the same formulation as in Section~\ref{Sec:System Model}. Replacing $\mathbf{y}[n]$ in Eq. \ref{Eq:EmpiricalCov} with $\widetilde{\mathbf{y}}_k[n]$ yields the windowed beamspace covariance matrix $\widetilde{\mathbf{R}}_k \in \mathbb{C}^{TW \times TW}$, which captures the spatial structure of the reduced beamspace signal. Using this matrix in Eq. \ref{Eq:CorrMVDR}, we obtain the reduced-dimension MVDR correlator:

\begin{equation}
\widetilde{\mathbf{c}}_k = \frac{ \widetilde{\mathbf{R}}_k^{-1} \, \widetilde{\mathbf{a}}_k }{ \widetilde{\mathbf{a}}_k^H \widetilde{\mathbf{R}}_k^{-1} \widetilde{\mathbf{a}}_k } \in \mathbb{C}^{TW},
\label{Eq:BeamCorrMVDR}
\end{equation}

where $\widetilde{\mathbf{a}}_k = \left\{\,\mathbf{S}_k\mathbf{D}\mathbf{a}^{(t)}_k\,\right\}_{t=1}^{T}$ is the windowed beamspace steering vector corresponding to the target direction \(k\).

To interpret the beamspace MVDR weights in the full antenna domain, we lift them back through the adjoint of the beamspace transform. Since the same beamspace transformation is applied to every tile, we first define the corresponding per-tile projection matrix
\begin{equation}
    \mathbf{B}_k \triangleq \mathbf{S}_k \mathbf{D} 
    \in \mathbb{C}^{W \times N},
    \label{eq:Pk_def}
\end{equation}

To apply this operator to the entire tiled array, we use a block-diagonal transformation expressed compactly through the Kronecker product with the identity matrix as

\begin{equation}
    \widetilde{\mathbf{y}}[n]
    = (\mathbf{I}_T \otimes \mathbf{B}_k)\mathbf{y}[n].
\end{equation}

Given the global beamspace MVDR weight vector 
$\tilde{\mathbf{c}}_k \in \mathbb{C}^{TW}$,
the corresponding lifted antenna-space correlator is defined as
\begin{equation}
    \widehat{\mathbf{c}}_k
    \triangleq
    (\mathbf{I}_T \otimes \mathbf{B}^H_k)\widetilde{\mathbf{c}}_k
    \in \mathbb{C}^{TN},
    \label{eq:lifted_ck}
\end{equation}
The lifted weight vector enables direct comparison with full-dimensional antenna-space beamformers and reveals the corresponding beam pattern and correlator structure in the original array domain.

\section{performance evaluation}
\label{Sec:performance evaluation}
\begin{figure}[t]
    \centering
    \includegraphics[width=0.75\linewidth, height = 0.5\linewidth]{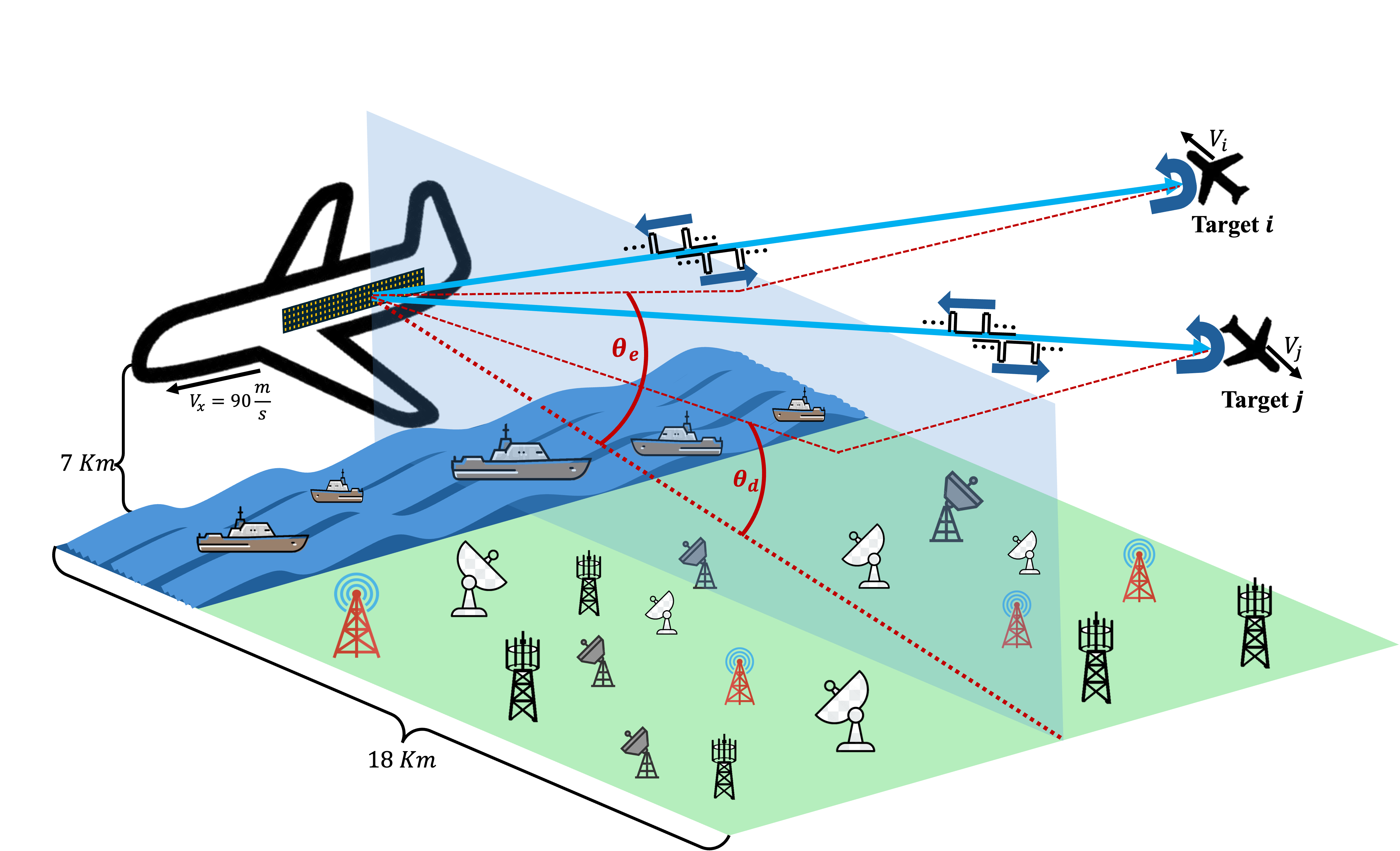}
    \caption{\small Simulated environment with airborne targets with ground- and sea-based interferers. Targets $i$ and $j$ are in Easy and Difficult modes, respectively.
    }
    \label{Fig:Environement}
\end{figure}

We evaluate the proposed tiled windowed--beamspace MVDR architecture using a simulated wideband radar dataset whose geometry, target ranges, and radial velocities are derived from the government-furnished data (GFD) for the DARPA SOAP (Scalable On-Array Processing) program. The spatial layout of the scene and the angular distribution of the targets follow those of the original SOAP dataset, ensuring that the simulated environment depicted in Fig.~\ref{Fig:Environement} reflects realistic radar operations.

While the SOAP GFD models interferers using a variety of communication and radar signals, we adopt a simplified wideband jammer model in which each interferer transmits spatially white Gaussian symbols with a specified power level relative to the targets. This enables controlled analysis of interference suppression under different interference-to-target power ratios.

The evaluation spans five scenarios, labeled A through E, with the number of interferers increasing from Scenario~A to Scenario~E. Each scenario is provided in two geometric configurations: an \textit{Easy mode} (index~1), in which targets exhibit greater elevation separation from the ground, and a \textit{Difficult mode} (index~2), in which this separation is reduced.

A full $16 \times 64$ (1024 element) array is partitioned into a $4 \times 2$ grid of tiles, each containing a $4 \times 32$ antenna subarray. The received wideband signal is channelized into $32$ subbands, which are processed independently using the coordinated tiled MVDR framework described in Section~\ref{Sec:BeamWindow}. To quantify the benefits of coordinated tiling, we compare the proposed architecture with a single-array windowed beamspace MVDR implementation. Target detection is performed using a 1D CFAR detector across the velocity bins of the range–Doppler map, with a threshold set at 10~dB above the estimated noise floor. Performance is evaluated in terms of range and velocity estimation error.

\subsection{Performance Benchmarks}
\label{Sec:Benchmarks}

We compare the detection performance of the proposed coordinated tiled windowed beamspace approach against single-tile windowed beamspace MVDR. Of particular interest is a comparison between a $2 \times 2$ beamspace window across 8 tiles against a $4 \times 8$ beamspace window for a single tile: both correspond to a 32-dimensional observation, and hence incur the same computational complexity for computing MVDR beamformers.  We also consider the incremental improvement from increased complexity within our architecture by considering a larger $4 \times 4$ per-tile beamspace window, corresponding to a 128-dimensional observation across 8 tiles.

To highlight the performance gap under mild and severe interference conditions, we present results for two representative scenarios: A1 (least severe) and E2 (most severe). In
Scenario~A1, the coordinated tiled processor maintains reliable detection performance even when the interference power is increased to more than 120~dB above the target power,
substantially exceeding the operating conditions of the original dataset. For Scenario~E2, which reflects the harshest geometry and interference placement, both methods are evaluated under interference levels up to 80~dB above the target power, matching the strongest interference conditions observed in the original dataset and applied uniformly across all interferers in our simulations.

For each target, we compute a detection SINR defined as the ratio between the peak power at the target’s range–velocity bin and the interference-plus-noise floor estimated via cell averaging in the surrounding CFAR stencil. This metric captures the quality of the
detection output, including robustness against false alarms and the ability to reliably set a detection threshold, while also reflecting the underlying interference suppression performance of the beamformer.

\subsection{Numerical Results}
\label{sec:Numerical_Results}

\begin{figure}[t]
    \centering
    \begin{subfigure}{0.48\textwidth}
        \centering
        \includegraphics[width=0.85\linewidth, height=0.4\linewidth]{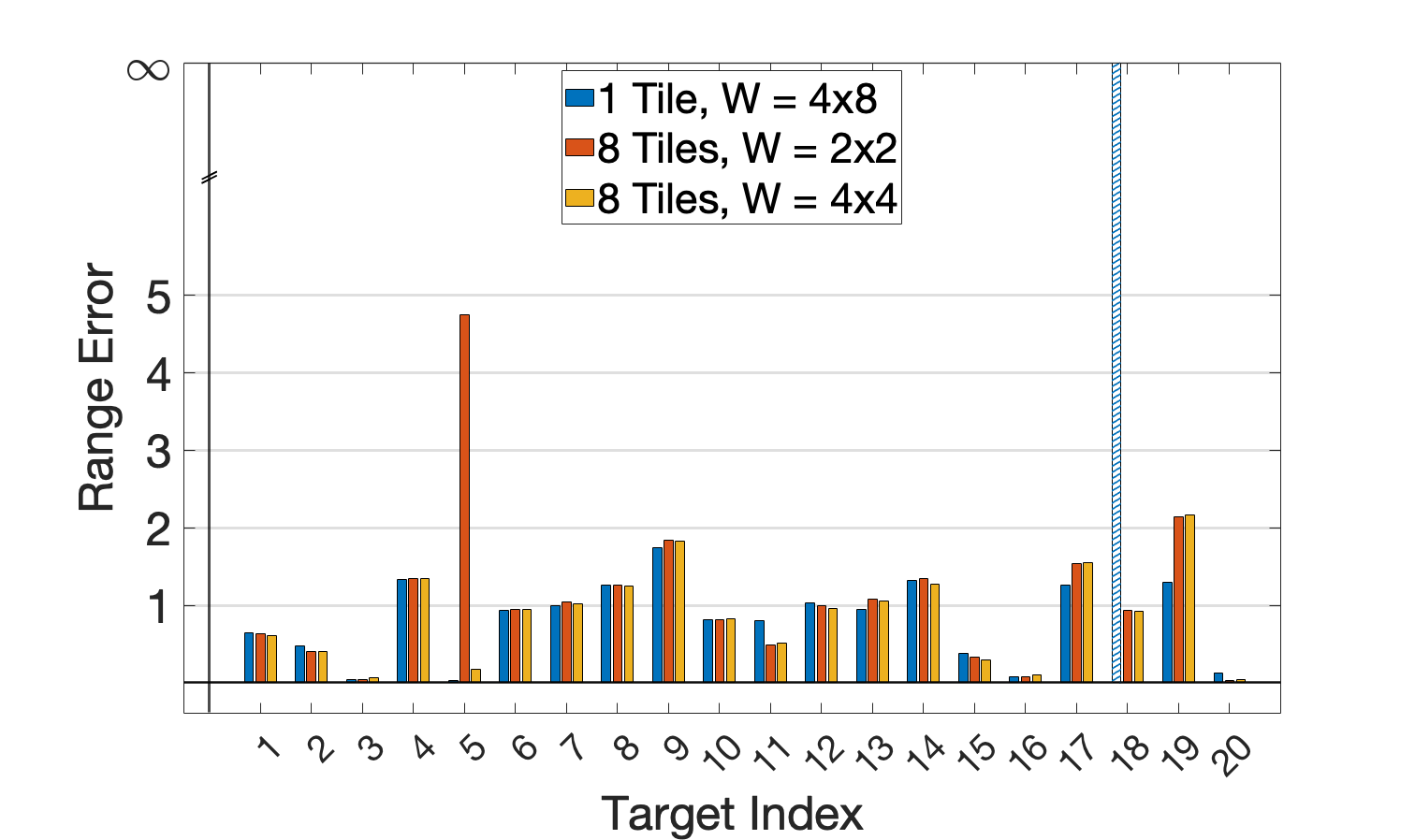}
        \caption{\footnotesize}
        \label{Fig:A1_Rng}
    \end{subfigure}
    \begin{subfigure}{0.48\textwidth}
        \centering
        \includegraphics[width=0.85\linewidth, height=0.4\linewidth]{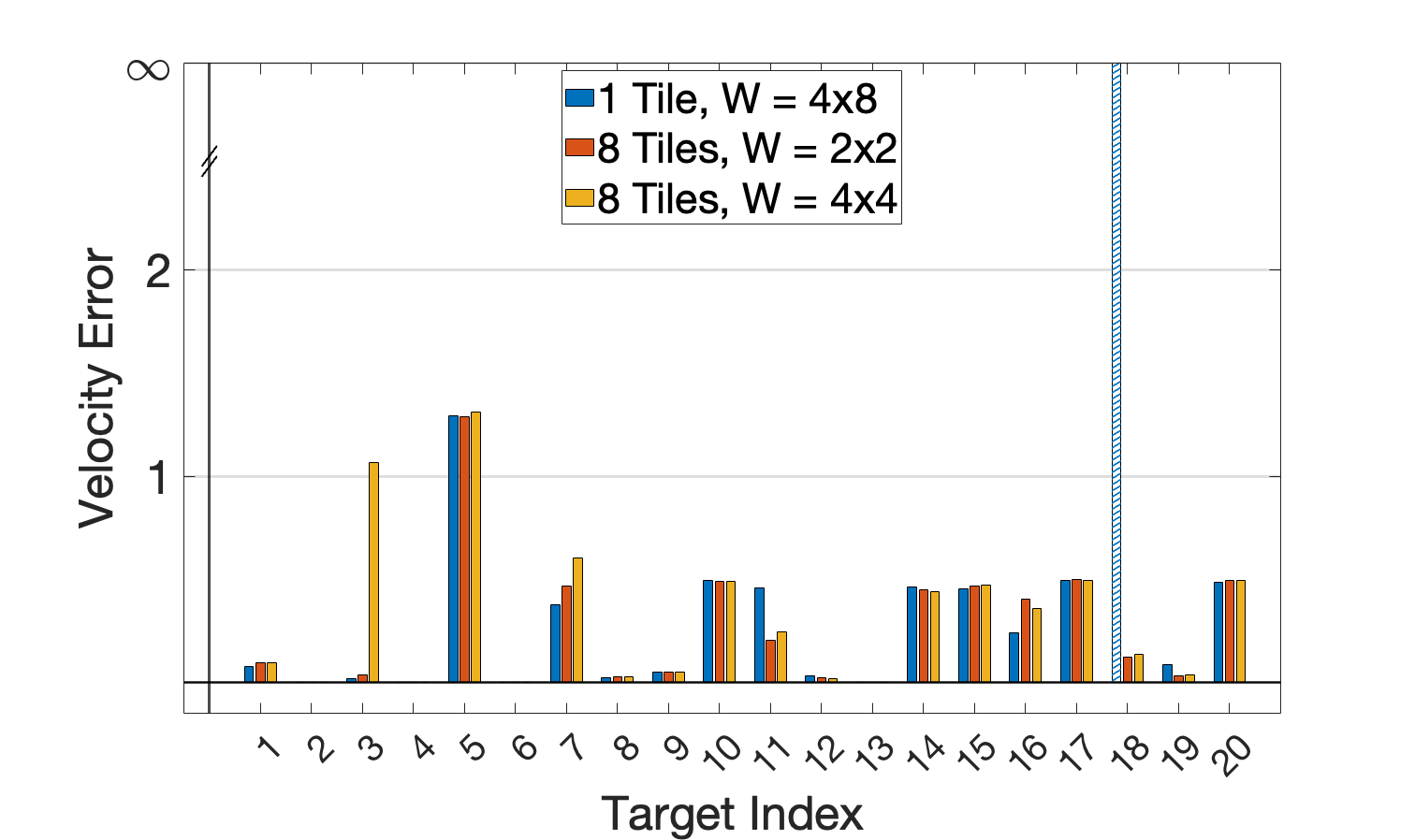}
        \caption{\footnotesize}
        \label{Fig:A1_Vel}
    \end{subfigure}
    \caption{\small Range and velocity estimation errors for Scenario~A1.}
    \label{Fig:A1RngVel}
\end{figure}
\begin{figure}[t]
    \centering
    \begin{subfigure}{0.48\textwidth}
        \centering
        \includegraphics[width=0.85\linewidth, height=0.45\linewidth]{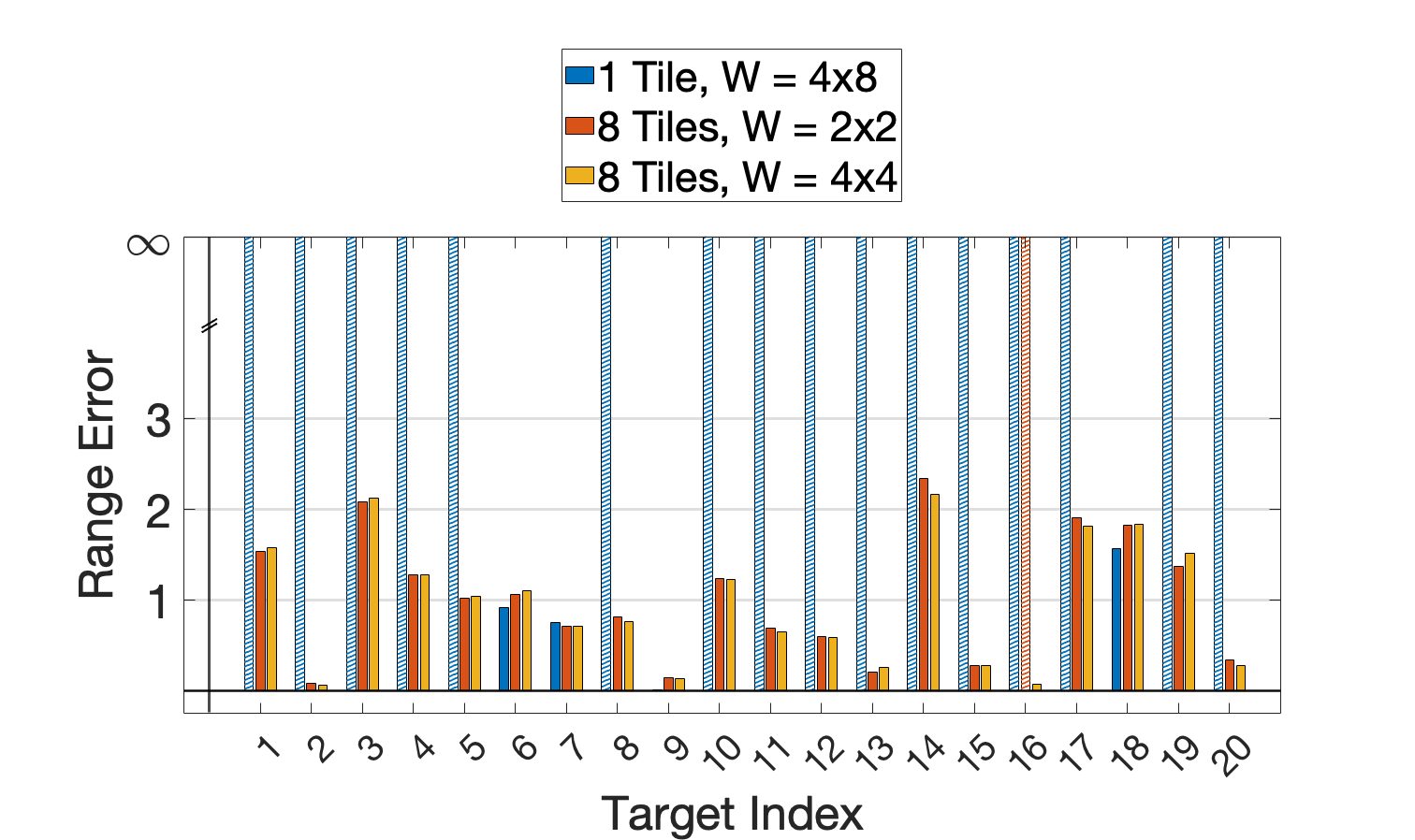}
        \caption{\footnotesize}
        \label{Fig:E2_Rng}
    \end{subfigure}
    \begin{subfigure}{0.48\textwidth}
        \centering
        \includegraphics[width=0.85\linewidth, height=0.45\linewidth]{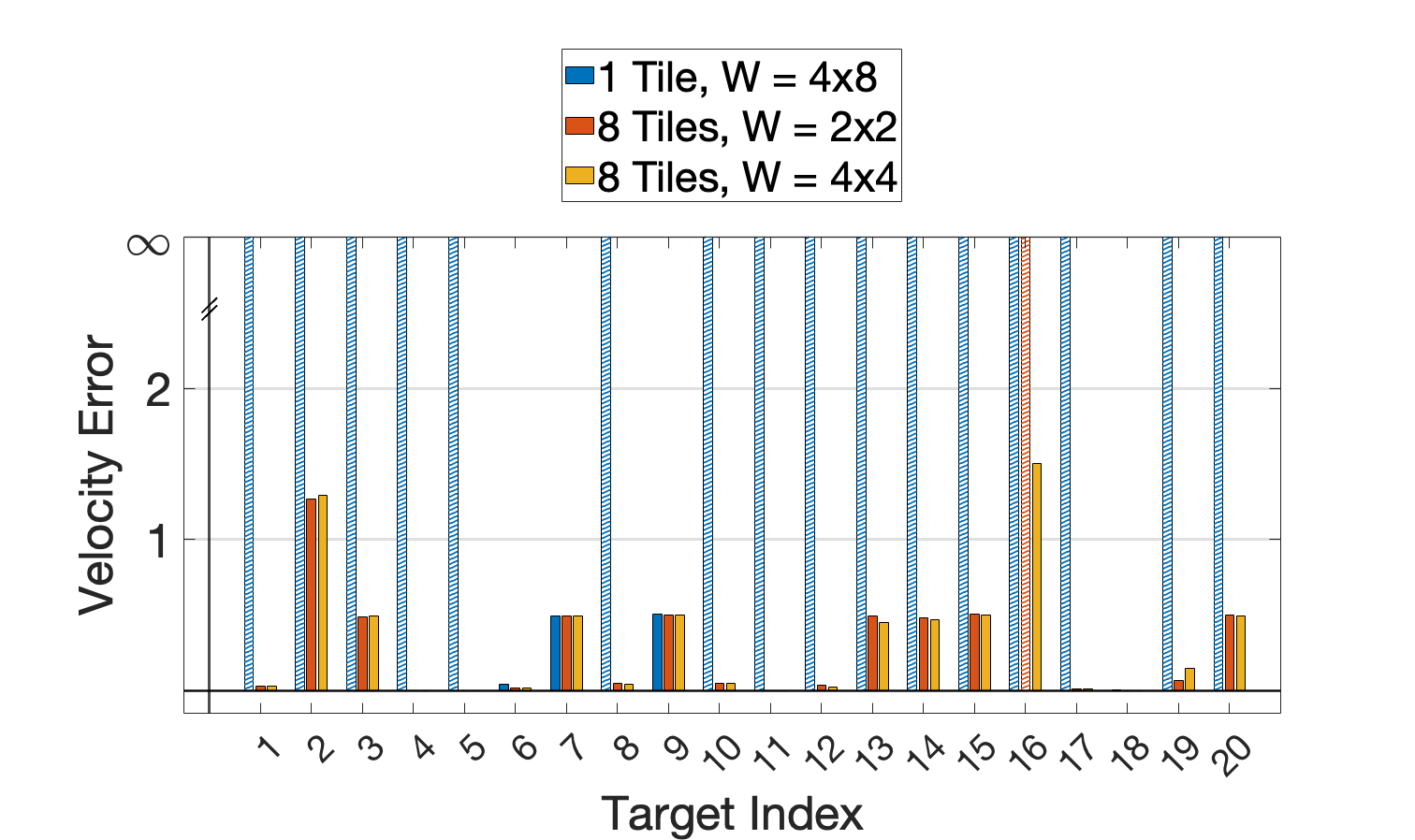}
        \caption{\footnotesize}
        \label{Fig:E2_Vel}
    \end{subfigure}
    \caption{\small Range and velocity estimation errors for Scenario~E2.}
    \label{Fig:E2RngVel}
\end{figure}

Figures~\ref{Fig:A1RngVel} and~\ref{Fig:E2RngVel} show the range and velocity estimation errors for the single-array beamspace MVDR and the proposed coordinated tiled MVDR architecture. In both scenarios, each tile operates with a beamspace window that is substantially smaller than its underlying antenna dimension.
As a result, even if the per-tile window size is kept equal to the window size used in the single-array processor, the total beamspace dimension in the tiled architecture remains
much smaller than that of full-dimensional MVDR. Coordinated processing across tiles further improves scalability: recovering full-array performance requires only a modest increase in per-tile window size, far smaller than the increase in total antenna count obtained by adding tiles. This demonstrates a key advantage of beamspace processing—its effective dimensionality grows slowly with array size due to energy concentration, allowing
the tiled system to outperform the single-array processor with significantly fewer beamspace coefficients.

Across both A1 and E2 scenarios, the coordinated tiled MVDR detects more targets and exhibits lower range and velocity errors for detected targets. Missed targets are shown in the figures using hatched bars, and their corresponding errors are depicted symbolically as infinity to indicate detection failure.

\begin{figure}[t]
    \centering
    \begin{subfigure}{0.48\textwidth}
        \centering
        \includegraphics[width=0.9\linewidth, height=0.5\linewidth]{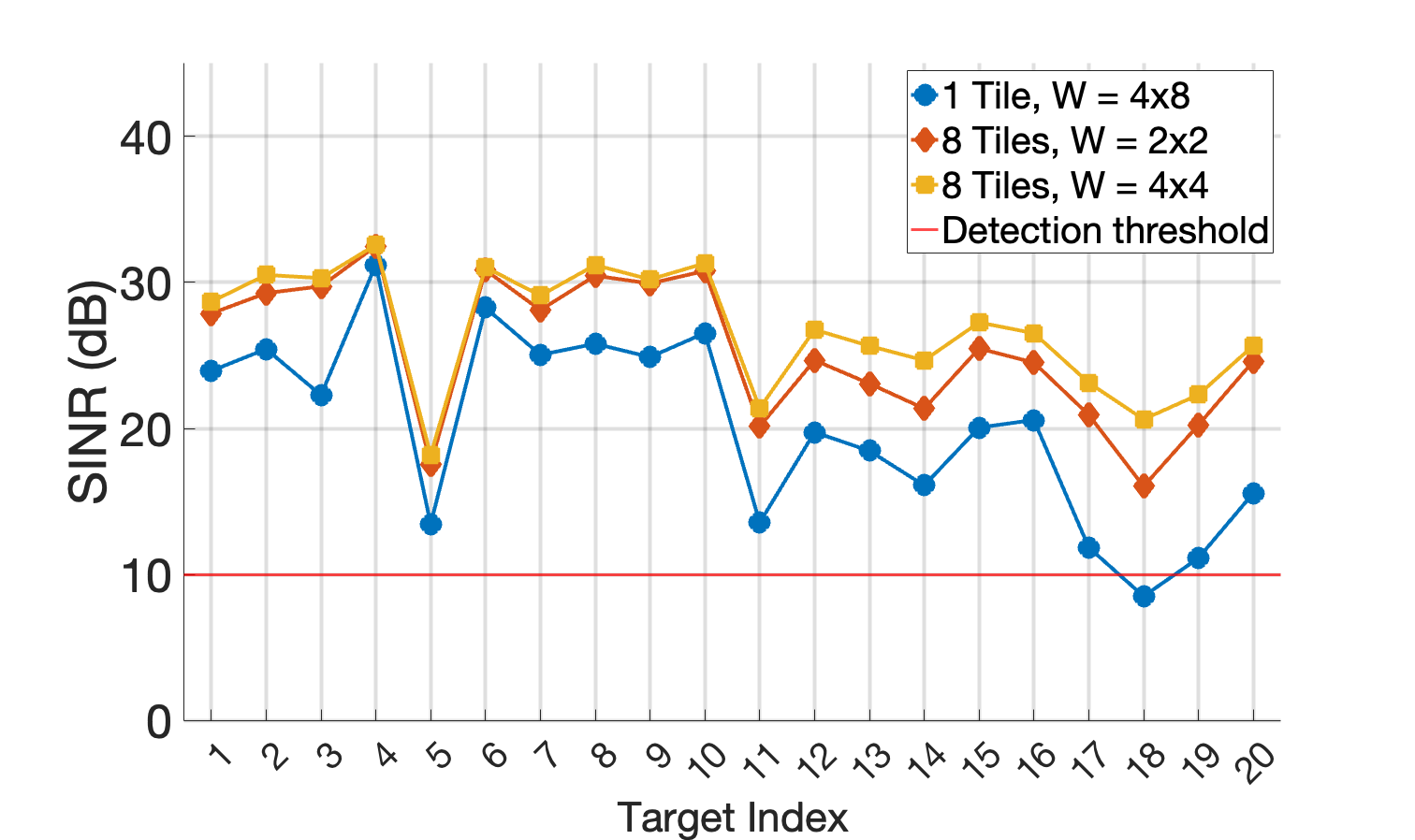}
        \caption{\footnotesize}
        \label{Fig:A1_SINR}
    \end{subfigure}
    \begin{subfigure}{0.48\textwidth}
        \centering
        \includegraphics[width=0.9\linewidth, height=0.5\linewidth]{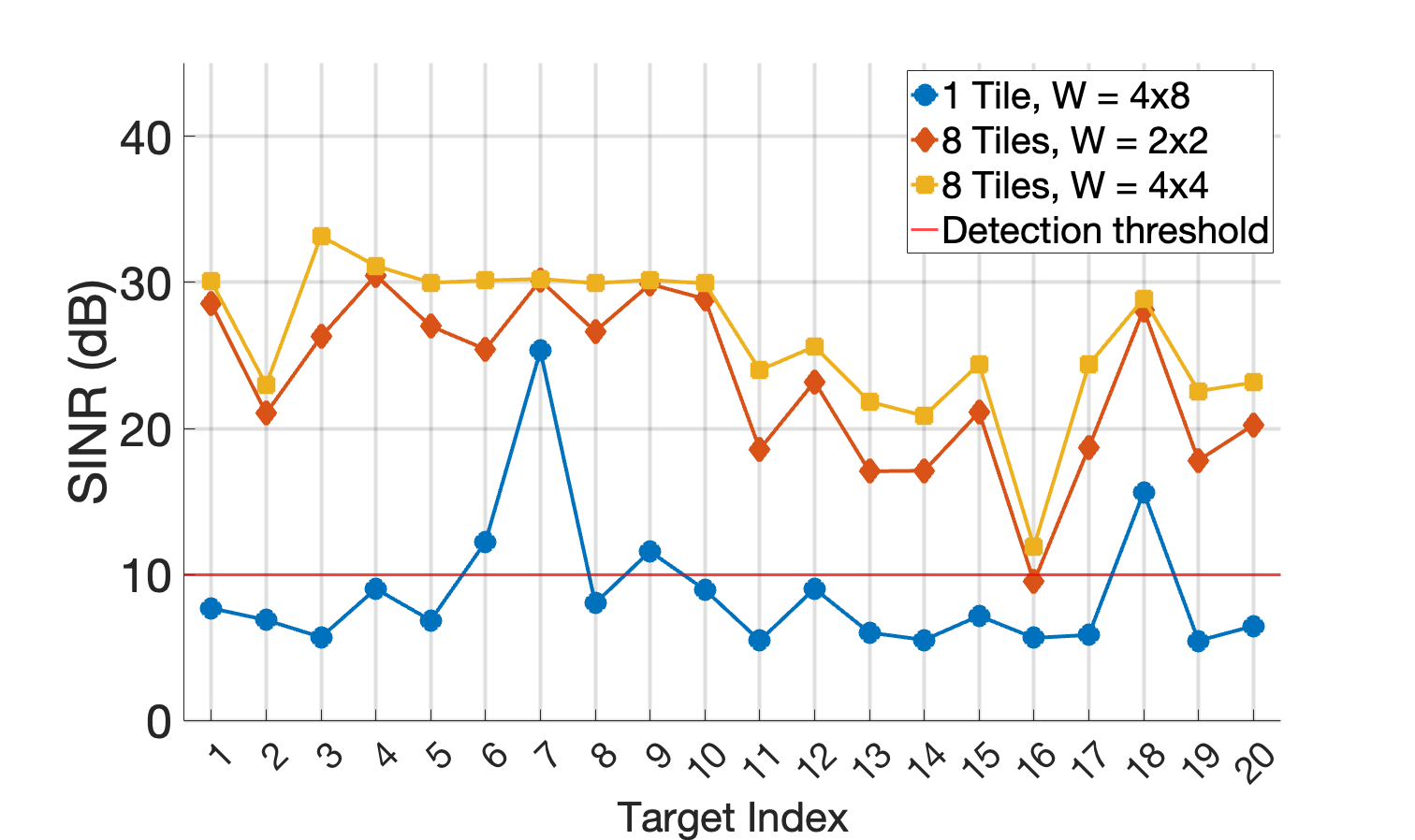}
        \caption{\footnotesize}
        \label{Fig:E2_SINR}
    \end{subfigure}
    \caption{\small SINR at the target grid points for both processing methods: (a)~Scenario~A1 and (b)~Scenario~E2.}
    \label{Fig:SINR}
\end{figure}

The SINR results shown in Figure~\ref{Fig:SINR} further highlight the advantage of the coordinated tiled MVDR architecture. Across both scenarios, the coordinated approach achieves consistently higher SINR at the target detection bins compared to the single-array processor. Targets with SINR below the detection threshold are considered missed detections.

The performance gap between the two methods becomes more pronounced in Scenario~E2, where the interference environment is significantly more severe. In this case, the coordinated
tiled processor maintains higher SINR for a larger fraction of the targets, demonstrating significantly increased robustness to strong interference than with a single tile.  

Using the lifted correlators, we can also characterize the effective beamforming pattern for target $k$ in the original antenna domain. For any azimuth–elevation pair $(\varphi,\theta)$, the normalized beam response is defined as
\begin{equation}
    P_k(\varphi,\theta)
    =
    \frac{
        \left| \langle \widehat{\mathbf{c}}_k , \mathbf{a}(\varphi,\theta) \rangle \right|
    }{
        \|\widehat{\mathbf{c}}_k\|_2 \, \|\mathbf{a}(\varphi,\theta)\|_2
    },
    \label{eq:cosine_similarity}
\end{equation}
which corresponds to the cosine similarity between the lifted correlator and the steering vector at $(\varphi,\theta)$.

\begin{figure}[t]
    \centering
    \begin{subfigure}{0.48\textwidth}
        \centering
        \includegraphics[width=\linewidth, height=0.5\linewidth]{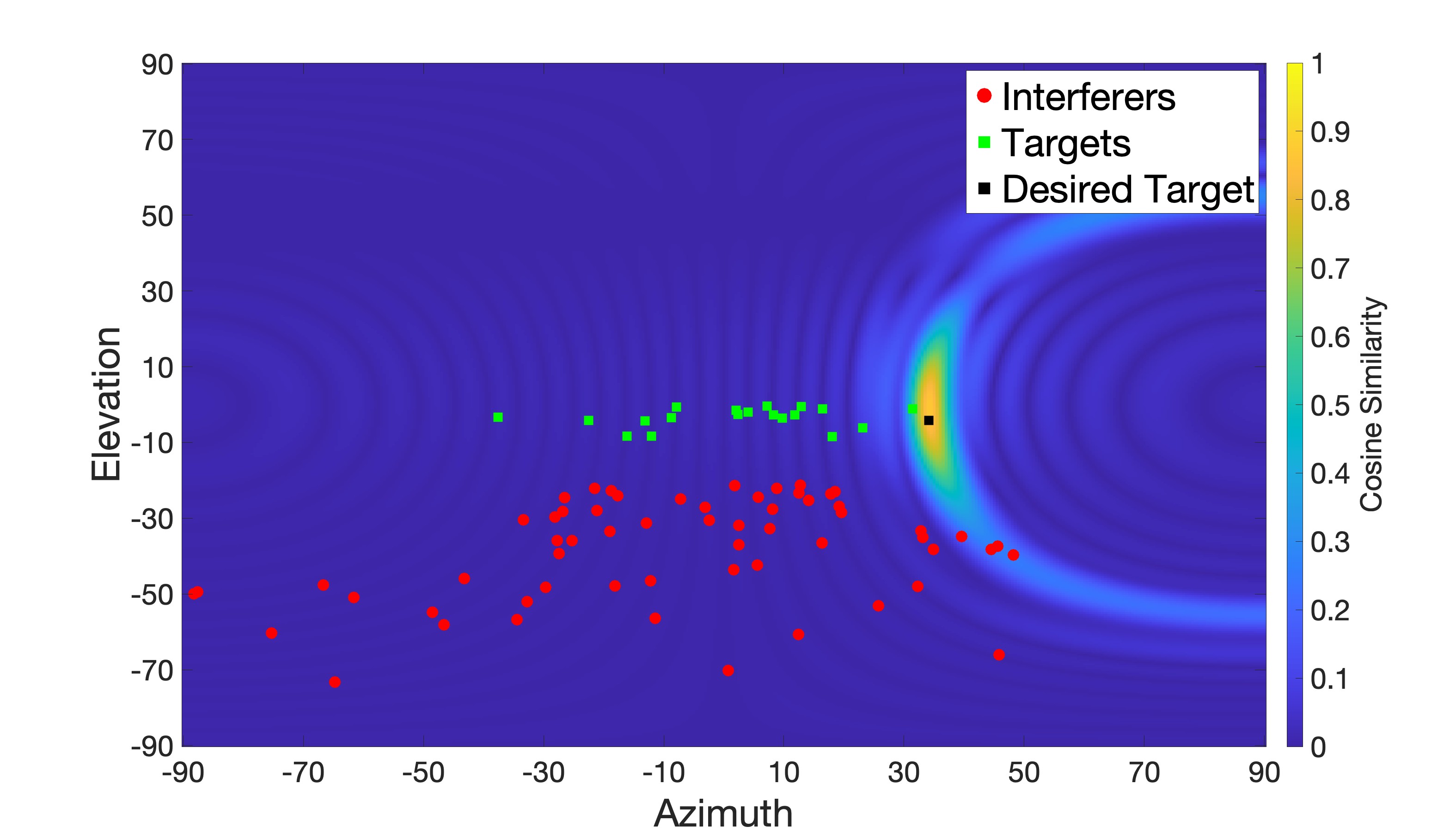}
        \caption{\footnotesize}
        \label{Fig:E2_SinglePattern}
    \end{subfigure}
    \begin{subfigure}{0.48\textwidth}
        \centering
        \includegraphics[width=\linewidth, height=0.5\linewidth]{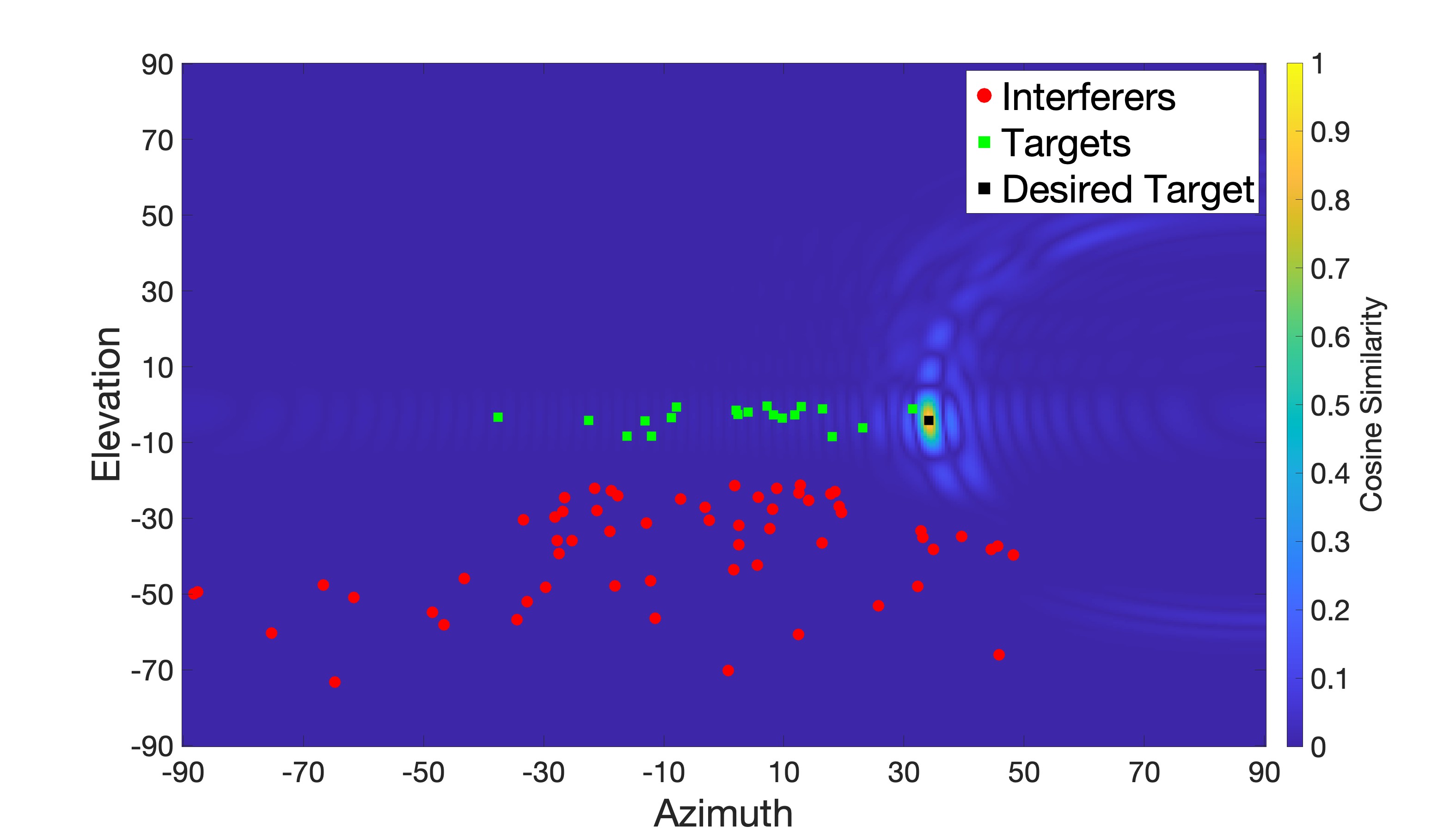}
        \caption{\footnotesize}
        \label{Fig:E2_TilePattern}
    \end{subfigure}
    \caption{\small Beamforming patterns for target~18 in ~$\text{E}_2$ and in the center subband:
    (a) single-array beamspace processing with a $4\times 8$ window,
    and (b) coordinated tiled beamspace processing with a $2\times 2$ window per tile.
    The tiled architecture achieves a narrower mainlobe and deeper nulls despite
    a significantly reduced per-tile dimension.}
    \label{Fig:BeamPattern}
\end{figure}

Figure~\ref{Fig:BeamPattern} compares this beamforming pattern for a representative target under (a) single-tile processing with a $4 \times 8$ beamspace window (i.e, a 32-dimensional observation) and (b) coordinated tiled processing with a $2 \times 2$ per-tile beamspace window (i.e., resulting in a 32-dimensional observation when aggregated over 8 tiles). The tiled configuration produces a narrower mainlobe, deeper spatial nulls, and less noise enhancement, explaining its improved detection performance. Thus, our coordinated tiled approach is able to exploit the 8-fold increase in antenna elements (relative to a single tile) to obtain a far superior beam pattern without increasing the computational complexity of beamforming.

\section{conclusion}
\label{Sec:conclusion}
The results in this paper demonstrate the promise of tiled beamspace architectures as a means of scaling the number of antenna elements and bandwidth for massive MIMO radar.  The key underlying concepts are energy concentration in beamspace, and parallelization of computation across tiles, subbands and targets. Ongoing research focuses on methods for further reduction in computational complexity and inter-tile communication, and on hardware-signal processing co-design for implementing the proposed architectures. While our performance evaluation focuses on beamforming towards targets being tracked, an important topic for future work is to develop efficient techniques for target acquisition. 


\section*{Acknowledgment}
This material is based upon work supported in part by the Defense Advanced Research Projects Agency (DARPA) under Agreement No. HR00112490508, and in part by the Center for Ubiquitous Connectivity (CUbiC), sponsored by Semiconductor Research Corporation (SRC) and DARPA under the JUMP 2.0 program.

\bibliographystyle{IEEEtran}
\bibliography{ref}

\end{document}